\documentclass[aps,prd,onecolumn,11pt,preprintnumbers,floats,floatfix,apsrev,amsmath,amsfonts,nofootinbib,superscriptaddress]{revtex4-2}
\usepackage{color}
\usepackage{graphicx}
\usepackage{dcolumn}
\usepackage{natbib}
\usepackage{hyperref}
\usepackage{hypernat}
\usepackage{natbib}
\usepackage{soul}
\usepackage{amssymb}
\usepackage{xcolor}
\usepackage{ulem}
\newcommand{\dd}{{\rm d}}

\newcommand{\p}{\partial}

\newcommand{\be}{\begin{equation}}
\newcommand{\en}{\end{equation}}

\newcommand{\paren}[1]{\left({#1}\right)}
\newcommand{\brac}[1]{\left[{#1}\right]}

\begin{document}
\title{Dynamical and thermodynamical stability of a charged thin-shell wormhole}

\author{Ernesto F. Eiroa}
\email{eiroa@iafe.uba.ar}

\author{Griselda Figueroa-Aguirre}
\email{gfigueroa@iafe.uba.ar}
\affiliation{Instituto de Astronom\'{i}a y F\'{i}sica del Espacio  (IAFE, CONICET-UBA),Ciudad Universitaria, 1428, Buenos Aires, Argentina.}

\author{Miguel~L.~Pe\~{n}afiel}
\email{miguelpenafiel@upb.edu}
\affiliation{UERJ - Universidade do Estado do Rio de Janeiro 150, CEP 20550-013, Rio de Janeiro, RJ, Brazil.}
\affiliation{FIA - Facultad de Ingeniería y Arquitectura, Universidad Privada Boliviana, Camino Achocalla Km 3.5, La Paz, Bolivia.}

\author{Santiago Esteban Perez Bergliaffa}
\email{santiagobergliaffa@uerj.br}
\affiliation{UERJ - Universidade do Estado do Rio de Janeiro 150, CEP 20550-013, Rio de Janeiro, RJ, Brazil.}

\date{\today}

\begin{abstract}
A study of the dynamical and thermodynamical stability of a charged thin-shell wormhole built by gluing two Reissner-Nordstr\"{o}m geometries is presented. The charge on the shell is linearly related to the matter content. For the dynamical stability, a concise inequality is obtained, valid for any barotropic equation of state that relates the pressure with the energy density at the throat. A thermodynamical description of the system is introduced, which leads to the temperature and the electric potentials. Adopting a linear equation of state for the pressure and a definite form for the entropy function, the set of equilibrium configurations that are both dynamically and thermodynamically stable is found.
\end{abstract}

\maketitle

\section{Introduction}

Wormholes (WHs) are solutions of a given gravitational theory that connect two regions of a single spacetime (or two spacetimes) through a throat (see \cite{Visser1995} for a review). In General Relativity, they are generated by exotic matter \cite{Morris1988} that should violate the null energy condition already at the classical level.\footnote{The total amount of exotic matter can be arbitrarily minimized under certain conditions, see for instance \cite{Visser2003}.}  
There are many ways to construct a wormhole, either in General Relativity (see for instance \cite{Lemos2003})
or in modified gravitational theories, such as Brans-Dicke theory \cite{Anchordoqui1996}, $f(R)$ theories \cite{Lobo2009}, the Randall-Sundrum model \cite{Anchordoqui1999}, modified teleparalell gravity \cite{Boehmer2012}, and hybrid metric-Palatini gravity \cite{Capozziello2012}, to mention just a few.
In addition to the consideration of WH solutions in different gravitational theories, such spacetimes can be constructed in a given gravitational theory either by using a continuous distribution of matter, or by restricting the matter content to a thin shell. In the latter approach,  pioneered by Visser \cite{Visser1989b}, it is possible to construct a thin-shell wormhole (TSW) by joining  two spacetimes by the ``cut and paste'' technique,  using the Darmois-Israel formalism \cite{Israel1966}. Among the many spherically symmetric TSWs obtained in this way, we can mention the one built by gluing of two Schwarzschild manifolds \cite{Visser1989a}, electrically charged wormholes in d-dimensional General Relativity with a cosmological constant \cite{Dias2010}, five dimensional geometries in Einstein-Maxwell theory with a Gauss-Bonnet term \cite{Thibeault2005}, spacetimes in dilaton gravity \cite{Eiroa2005} and in $f(R)$ theories \cite{Eiroa2015h,Eiroa2016z,Lobo2020v,Habib2020r}, and also cylindrical TSWs \cite{Eiroa2004b, Mazha2014}.\footnote{For a review, see \cite{Lemos2021b}.}

The study of dynamical stability of spherically symmetric TSWs through the analysis of the linearized perturbations of the radius of the throat was presented by Poisson and Visser in \cite{Poisson1995}, extended to include the presence of a cosmological constant \cite{Lobo2003}, to charged TSWs constructed by joining two Reissner-Nordstr\"{o}m spacetimes in \cite{Eiroa2004}, and generalized in \cite{Eiroa2008k, Garcia2011}.\footnote{The linear  stability of TSW when perturbed away from a evolving initial state was presented in \cite{Li2018}.} 
It was shown in \cite{Eiroa2004} that overcharged TSWs possess larger stability regions in the corresponding parameter space than TSWs for which the ADM mass is larger than the charge. 
For the linear stability of a rotating wormhole built by joining two Kerr solutions, see {\cite{Sharif2020}.

An important feature of thin-shell spacetimes is the fact that they allow a consistent thermodynamical description of the matter on the shell, thus paving the way to the examination of the thermodynamical stability of these solutions. Thermodynamical stability assures that no inhomogeneities that may eventually lead to a phase transition are formed \cite{Callen1985}. The entropy and thermodynamical stability of an uncharged spherical shell joining Minkowski spacetime with Schwarzschild geometry was first studied by Martinez \cite{Martinez1996}. Later, Lemos \textit{et al.} \cite{Lemos2015} calculated the entropy of a self-gravitating electrically charged thin shell
separating flat spacetime from Reissner-Nordstr\"{o}m solution (see \cite{Lemos2015a} for the extremal case). 
The parameter space that corresponds to configurations with a non-charged shell that are both dynamically and thermodynamically stable was studied in \cite{Bergliaffa2020}. This analysis was later extended to the charged case in \cite{Reyes2022}. The results in the last two references show that the  imposition of both types of stability may lead to strong constraints on the parameter space of the system.

Just as in the case of spherical thin-shells, it should be possible to derive a thermodynamical description for 
the matter sourcing a TSW and also, in the process, to analyze the thermodynamical stability of these solutions, as well as their dynamical stability.  Such an analysis was presented in \cite{Forghani2019}, for an  uncharged TSW constructed {by joining two copies of Schwarzschild spacetime.} In the present work, we analyze the thermodynamical stability of charged TSWs and study the different situations in which both dynamical and thermodynamical stability conditions are simultaneously satisfied.
The article is organized as follows: in Sect. \ref{sec:TSWs} we present the procedure of construction of a TSW and obtain the relevant physical quantities.  The dynamical stability of the charged TSW is studied in Sect. \ref{sec:dynstability}, introducing a formalism that reduces the problem to a single inequality for two parameters of the TSW, and re-obtaining in the process some of the previous results found by Eiroa and Romero in \cite{Eiroa2004}. A thermodynamical description for the shell is developed in Sect. \ref{sec:thstab}, assuming that its entropy is described by three extensive variables, namely, mass, charge, and area. 
The regions of the parameter space for which the configurations are \textit{completely stable} (\textit{i.e.} both dynamically and thermodynamically stable) are presented in Sect. \ref{condsstab}, using the well-known inequalities involving the second derivatives of the entropy \cite{Callen1985}. We close with a discussion and some possible extensions of this work in Sect. \ref{sec:conclusions}.

Throughout this paper we adopt geometrized units where $G=c=1$, and the signature $\paren{-,+,+,+}$.

\section{Charged thin-shell wormholes}\label{sec:TSWs}

The Reissner-Nordstr\"{o}m (RN) metric is given by
\be \label{eq:dsRN}
\dd s^2=-\paren{1-\frac{2m}{r}+\frac{Q^2}{r^2}}\dd t^2+\paren{1-\frac{2m}{r}+\frac{Q^2}{r^2}}^{-1}\dd r^2+r^2\dd\Omega^2\ ,
\en
where $m$ is the ADM mass, $Q$ is the electric charge, and $d\Omega^2\equiv\dd\theta^2+\sin^2\theta\dd\varphi^2$. If $Q\neq0$ a radial electric field is present, pointing outwards for $Q>0$ and inwards for $Q<0$. When $Q=0$, the Schwarzschild spacetime is recovered. As the sign of $Q$ only affects the  direction in which the electric field points, which is irrelevant for our study, from now on we adopt $Q \ge 0$ without losing generality.

In the black hole case, the metric above has two horizons, given by the roots of ${g_{00}}=0$, namely,
\be
r_{\pm}=m\pm\sqrt{m^2-Q^2}\ ,
\en
where $r_+$ is the (outer) event horizon and $r_-$ is the (inner) Cauchy horizon. Both horizons coincide if the condition $m=Q$ is satisfied. In such a case the corresponding spacetime is said to be extremal. 

The construction of a charged TSW is based on gluing two identical regions of the RN spacetime, defined by 
\be
\mathcal{M}^{\pm}=\{x/r\ge a\}.
\en
In particular, we will assume that $a>r_+$ in the case with $m\ge Q$,\footnote{The interesting case in which the throat of the wormhole is inside the event horizon has been considered, for a model that does not use thin shells, in \cite{Nojiri2024}.} and that $a>0$ for $m<Q$. The regions $\mathcal{M}^{\pm}$ are glued  at the timelike hypersurface $\Sigma=\Sigma^{\pm}=\{x/r-a=0\}$. 
The resulting manifold $\mathcal{M}=\mathcal{M}^+\cup\mathcal{M}^-$ is geodesically complete and corresponds to a TSW, with a throat of radius $a$ located at the shell. 

The metric $h_{ab}$, defined  on $\Sigma$, \textit{i.e.} for $r=a$, is that of a 2-sphere with the addition of a time coordinate, and can be written as
\begin{align}\label{RN9}
 ds_{\Sigma}^2 &=h_{ab}dy^a dy^b=-d\tau^2 + a^2(\tau)d\Omega ^2,
\end{align}
where 
$y^a=(\tau,\theta,\phi)$
and $\tau$ is the proper time for an observer located on the shell. 

The Darmois-Israel formalism requires the induced metric $h_{ab}$ to be continuous on the shell, and the discontinuity in the extrinsic curvature to be proportional to the stress-energy tensor of the matter on the shell.
Let us denote both sides of $\Sigma $ by $+$ and $-$. Therefore, the extrinsic curvature on each side of the hypersurface is defined as
\be \label{eq:extrinsic}
K^{a}_{\pm\  b}=\paren{\nabla_\alpha n^{\pm}_\beta}e^{\alpha}_{\pm \ c}e^{\beta}_{\pm\ b}h^{ac}\ ,
\en
where $n^{\pm}_\alpha$ is the vector normal to $\Sigma $ on each side, respectively, and $e^{\alpha}_{\pm\ a}$ is the vector tangent to it. The relevant components of the extrinsic curvature are
\begin{align}
K^{\tau}_{\pm\ \tau}&=\pm\frac{\frac{m}{a^2}-\frac{Q^2}{a^3}+\ddot{a}}{\sqrt{1-\frac{2m}{a}+\frac{Q^2}{a^2}+\dot{a}^2}}\ , \\
K^{\theta}_{\pm \ \theta}&=K^{\varphi}_{\pm\ \varphi}=\pm\frac{1}{a}\sqrt{1-\frac{2m}{a}+\frac{Q^2}{a^2}+\dot{a}^2}\ ,
\end{align}
where the overdot denotes derivative with respect to the proper time $\tau$.
They are connected to the stress-energy tensor on the shell $S_{ab}$, through the Lanczos equation
\be\label{eq:Lanczos}
S_{ab}=-\frac{1}{8\pi}\paren{\left[K_{ab}\right]-h_{ab}\left[K\right]}\ ,
\en
where the jump in the extrinsic curvature is given by $\left[K_{ab}\right]=K^{+}_{ab}-K^{-}_{ab}$.

For a perfect fluid, this tensor has the form $S^{a}_{\;b}=\text{diag}\paren{-\sigma,p,p}$, with $\sigma$ the surface energy density and $p$ the tangential pressure of the fluid, then from Eq. \eqref{eq:Lanczos} one has
\begin{align} 
\sigma&=-\frac{1}{2\pi a}\sqrt{1-\frac{2m}{a}+\frac{Q^2}{a^2}+\dot{a}^2}\ , \label{eq:sigma} \\
p&=\frac{1}{4\pi a}\frac{1-\frac{m}{a}+\dot{a}^2+a\ddot{a}}{\sqrt{1-\frac{2m}{a}+\frac{Q^2}{a^2}+\dot{a}^2}}  \label{eq:p}\ .
\end{align}
Taking into account that $A=4\pi a^2$ is the area of the WH throat, from Eqs. \eqref{eq:sigma} and \eqref{eq:p} the energy-momentum conservation takes the form
\be
\frac{d}{d\tau}\paren{\sigma A}+p\frac{dA}{d\tau}=0\ ,
\label{eq:cons1}
\en
which can also be written 
as 
\be
\dot{\sigma}=-2\paren{\sigma+p}\frac{\dot{a}}{a}\ .
\label{eq:cons2}
\en
For a given equation of state (EOS), the latter equation can be integrated, leading to $\sigma=\sigma(a)$.

A radial perturbation of an equilibrium configuration with $a=a_0$ causes $a$, $\sigma$, and $p$ to become functions  of $\tau$. Assuming a particular EOS in Eq. \eqref{eq:cons2} to obtain $\sigma=\sigma(a)$, it follows from Eq. \eqref{eq:sigma} that the differential equation
\be \label{eq:evoTSW}
\dot{a}^2-\frac{2m}{a}+\frac{Q^2}{a^2}-4\pi^2a^2\sigma^2=-1 
\en
determines the dynamics of the throat.

\section{Dynamical stability} \label{sec:dynstability}

The general framework for the study of the linearized stability of TSWs was set up by Poisson and Visser in \cite{Poisson1995}. Following this approach, Eiroa and Romero \cite{Eiroa2004} analyzed the linearized dynamical stability of the charged TSW.\footnote{A comparison of the dynamical stability of a shell joining Minkowski and RN spacetimes and thin-shell charged wormholes was presented in \cite{Sharif}.} 
In this section, we will review some of these results and implement a formalism to present them in a form that will be useful in the analysis of the thermodynamical stability.

The dynamical equation \eqref{eq:evoTSW} can be written as
\be \label{eq:dyneq}
\dot{a}^2+V(a)=0\ ,
\en
where
\be \label{eq:pot}
V(a)\equiv 1-\frac{2m}{a}+\frac{Q^2}{a^2}-\left[ 2\pi a\sigma(a)\right] ^2\ .
\en
We will assume perturbations around a static solution with $a=a_0$. From Eqs. \eqref{eq:sigma} and \eqref{eq:p} the energy density and the pressure are given by
\begin{align}
\sigma_0&=-\frac{1}{2\pi a_0}\sqrt{1-\frac{2m}{a_0}+\frac{Q^2}{a_0^2}}\ , \label{eq:sigma0}\\
p_0&=\frac{1}{4\pi a_0}\frac{1-\frac{m}{a_0}}{\sqrt{1-\frac{2m}{a_0}+\frac{Q^2}{a_0^2}}} \label{eq:p0}\ .
\end{align}

The material mass is defined as {$M=4\pi a_0^2 \sigma_0$}. Since from Eq. \eqref{eq:sigma0} it is seen that the energy density is negative, the total material mass is also negative. 
By virtue of Eq. \eqref{eq:sigma0}, the ADM mass can be written in terms of the material mass as
\be \label{eq:mADM}
m=\frac{a_0}{2}-\frac{M^2}{8a_0}+\frac{Q^2}{2a_0}\ .
\en
The features of the static TSW in terms of $M$, $a_0$, and $Q$ are  determined by Eqs. \eqref{eq:sigma0}-\eqref{eq:mADM}.
Using Eq. \eqref{eq:mADM}, the potential in Eq. \eqref{eq:pot} 
can be rewritten as
\be
V(a) = 1 -\frac{a_0}{a} +\frac{M^2}{4a a_0} +\left(
\frac{1}{a}-\frac{1}{a_0}\right) \frac{Q^2}{a} -\left( 2\pi a \sigma
\right) ^2\ ,
\en
which, when evaluated at $a=a_0$ takes the simple form
\be
V(a_0)=\frac{M^2}{4 a_0^2}-4 \pi ^2 a_0^2 \sigma_0 ^2\ ,
\en
while the energy density and the pressure are given by
\be
\sigma_0=\frac{M}{4\pi a_0^2}\ ,
\en
\be
p_0=-\frac{4a_0^2+M^2-4Q^2}{16\pi a_0^2 M}\ ;
\en
thus, it is easy to verify that $V(a_0)=0$.

Defining the superficial charge density $\sigma_e $ by $Q=4\pi a^2\sigma_e$; then, the radius of the throat, the material mass, and the ADM mass can be written as follows:
\begin{align}
a_0&=\frac{1}{2\pi\sqrt{4\paren{\sigma_e^2-p_0\sigma_0}-\sigma_0^2}} \ , \\
M&=\frac{\sigma_0}{\pi\brac{4\paren{\sigma_e^2-p_0\sigma_0}-\sigma_0^2}}\ , \\
m&=\frac{4\sigma_e^2-2p_0\sigma_0-\sigma_0^2}{2\pi\brac{4\paren{\sigma_e^2-p_0\sigma_0}-\sigma_0^2}^{3/2}}\ .
\end{align}
By assuming a linear relation between the matter density and the charge density, namely 
\be\label{eq:beta}
\sigma_e=\beta\vert\sigma_0\vert\ ,
\en
as well as a barotropic EOS, given by $p=x(\sigma)\sigma$, 
it follows that
\begin{align}
a_0&=\frac{1}{2\pi\vert\sigma_0\vert\sqrt{4\paren{\beta^2-x}-1}}\ , \label{eq:ax} \\
M&=\frac{1}{\pi\sigma_0\brac{4\paren{\beta^2-x}-1}}\ , \label{eq:Mx}\\
m&=\frac{4\beta^2-2x-1}{2\pi\vert\sigma_0\vert\brac{4\paren{\beta^2-x}-1}^{3/2}}\label{eq:mx}\  .
\end{align}
Note that $\beta$ and $Q$ have the same sign, so our choice of $Q \ge 0$ results in $\beta \ge 0$ without losing generality.

In order to avoid non-real values for $a_0$ and $m$, the bound 
\be\label{eq:physcond1}
4\paren{\beta^2-x}-1>0\ 
\en
should be fulfilled. It is still necessary to impose that the ADM mass to be non-negative. From Eq. \eqref{eq:mx}, it follows that
\be \label{eq:physcond2}
4\beta^2-2x-1\ge0
\en
should also be satisfied.

The stability analysis is  based on the expansion of the potential \eqref{eq:pot} around the static solution, given by
\be\label{eq:TaylorV}
V(a)=V(a_0)+V'(a_0)\paren{a-a_0}+\frac{V''(a_0)}{2}\paren{a-a_0}^2+\mathcal{O}\paren{a-a_0}^3\ .
\en
By differentiating Eq. \eqref{eq:pot}, it follows that
\begin{align}
V'(a)&=\frac{2m}{a^2}-\frac{2Q^2}{a^3}-8\pi^2a\sigma\paren{a\sigma'+\sigma} \ ,\\
V''(a)&=-\frac{4m}{a^3}+\frac{6Q^2}{a^4}-8\pi^2\paren{4a\sigma\sigma'+\sigma^2+a^2\sigma'^2+a^2\sigma\sigma''}  \label{eq:Vpp1}\ .
\end{align}
Since $\sigma'=\dot{\sigma}/\dot{a}=-2\paren{\sigma+p}/a$ 
and 
$\sigma''=-2(\sigma'+p')/a+2(\sigma+p)/a^2$,
these expressions can be rewritten in the form
\be
V'(a)=\frac{2m}{a^2}-\frac{2Q^2}{a^3}+8\pi^2 a\sigma\paren{2p+\sigma}\ ,
\en
\be
V''(a)=-\frac{4m}{a^3}+\frac{6Q^2}{a^4}-8\pi^2\left[\paren{\sigma+2p}^2+2\sigma\paren{\sigma+p}\paren{1+2\frac{p'}{\sigma'}}\right] \label{eq:Vppa}.
\en
Using the derivatives of the potential evaluated at the static solution $a_0$ in the series expansion for $V(a)$, and taking into account that $V(a_0)=V'(a_0)=0$, it follows that
\be
V(a)=\frac{V''(a_0)}{2}\paren{a-a_0}^2+\mathcal{O}\paren{a-a_0}^3\ ,
\en
where the second derivative reads
\be
V''(a_0)=-\frac{2}{a_0^4} \left\{\frac{a_0 \left[(a_0-m)^3+m \left(m^2-Q^2\right)\right]}{a_0^2-2 a_0 m+Q^2}
+2  \left(a_0^2-3 a_0 m+2 Q^2\right)\frac{p'}{\sigma'}\right\} \ .\label{eq:Vppa0}
\en
This expression is valid for any barotropic EOS. Note that Eq. \eqref{eq:dyneq} implies that only values of the parameters for which $V(a)\leq 0$ are allowed. For a given $a_0$, a sufficient condition for stability is $V''(a_0)>0$. In \cite{Eiroa2004}, Eiroa and Romero studied the case of a constant ratio $p'/\sigma'$ and the consequent stability of the configurations. In the present work, we will first examine this particular scenario before dealing with the general case. 

\subsection{Dynamical stability for a linear equation of state}\label{dinstab}

Let us examine the case in which $x$ is a constant, denoted by $\kappa$. From Eqs. \eqref{eq:ax}-\eqref{eq:mx}, the stability condition for a linear EOS  is determined by the inequality
\be \label{eq:stabsimple}
V''(a_0)=8 \pi ^2 \sigma_0 ^2 \left(4 \beta ^2-8 \kappa^2-6 \kappa-1\right)>0\ ,
\en
where $\sigma_0$ in terms of $a_0$, $\beta$, and $\kappa $ reads 
\be \label{sigma-beta-kappa}
\sigma_0= -\frac{1}{2\pi a_0 \sqrt{4(\beta ^2-\kappa)-1}}. 
\en
As a consequence that $\sigma_0$ is squared in Eq. \eqref{eq:stabsimple}, the dynamical stability is independent of $\sigma_0$, and therefore also of the throat radius $a_0$. The region satisfying this condition in the $(\beta, \kappa)$ space is given by
\be
-\frac{3}{8}-\frac{1}{8}\sqrt{1+32\beta^2}<\kappa <-\frac{3}{8}+\frac{1}{8}\sqrt{1+32\beta^2}
\en
and displayed in Fig. \ref{fig:dyn_stab}. A small part of this region is discarded from our analysis, because it falls within the non-physical zone where the conditions stated by  Eqs. \eqref{eq:physcond1} and \eqref{eq:physcond2} are not fulfilled. It can be seen from the plot that the region of dinamically stable TSWs is inside the one corresponding to the overcharged case for which the relation $m< Q$ holds, that is
\be
-\beta-\frac{1}{2}<\kappa<\beta-\frac{1}{2}.
\label{overQ}
\en

\begin{figure}[t!]
\centering
\includegraphics[width=0.4\textwidth]{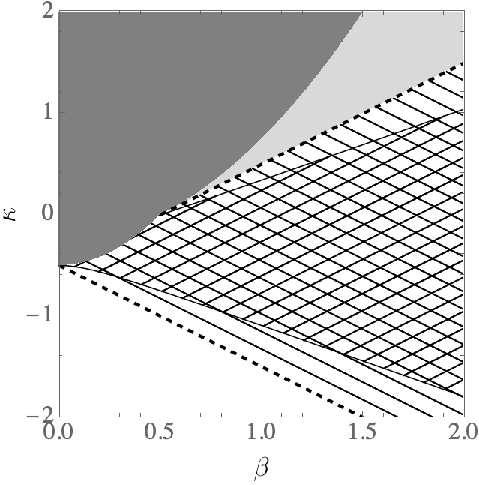}
\caption{The dynamical stability for the EOS $p=\kappa\sigma$, $\kappa\in\mathbb{R}$, is represented by the upward-line meshed region in the $(\beta,\kappa)$ plane. The downward-line meshed region corresponds to the overcharged case, bounded by the dashed lines $ \kappa = \pm \beta -1/2$ for which  $Q=m$. Configurations in the light gray zone have $a_0<r_+$ and are not considered in this work.  The medium gray region is non-physical (see text). }
\label{fig:dyn_stab}
\end{figure}

Let us remark that, although $\kappa$ is the squared speed of sound in the case of ordinary matter, for which $0<\kappa\le 1$,  this may not be the case for exotic matter \cite{Poisson1995}. In the absence of a microphysical description of the latter, we consider in Fig. \ref{fig:dyn_stab} the possibility of having any real value of $\kappa $.

\subsection{A more general equation of state}

Let us now assume that $p=x(\sigma) \sigma$, with $x(\sigma)$ an arbitrary function. 
Then, it follows that
\be
\frac{dp}{d\sigma}=\frac{dx(\sigma)}{d\sigma}\sigma + x(\sigma)\ .
\nonumber
\en 
From Eq. \eqref{eq:Vppa},  taking into account Eqs. \eqref{eq:ax}-\eqref{eq:mx}, the second derivative of the potential is actually a function of $\sigma$, $\beta$, and $x(\sigma)$,  which evaluated at $a_0$ has the form
\be
V''(a_0)= 8 \pi ^2 \sigma_0^2 \left[ 4\beta ^2 -8x_0^2
-6x_0-1-4( 1+x_0)\sigma_0\left. \frac{dx(\sigma)}{d\sigma}\right|_{a_0} \right],
\en
where $x_0\equiv x(\sigma)|_{a_0}$. In order to determine the regions of the parameter space in which  $V''(a_0)>0$, it is necessary to choose a  function $x=x(\sigma)$.

\section{Thermodynamics of a charged TSW}
\label{sec:thstab}

Let us introduce the equations that define the functions associated with the thermodynamics of the object of interest. We will mainly follow the approach presented in \cite{Lemos2015}, with some differences due to the particularities of the system studied in the present work.

Assuming that the shell is in equilibrium with a well-defined temperature $T(M,A,Q)$ and entropy $S(M,A,Q)$, the first law of thermodynamics is
\be\label{eq:firstlaw}
TdS=dM+pdA-\Phi dQ\ ,
\en
where $\Phi$ is the electric potential on the shell. Defining $\alpha\equiv1/T$, the latter equation can be written as
\be\label{eq:firstlawalpha}
dS=\alpha dM+\alpha pdA-\alpha\Phi dQ\ .
\en
In order for the differential $dS$ to be exact, the following integrability conditions should be satisfied
\begin{align}
\paren{\frac{\p\alpha p}{\p M}}_{A,Q}&=\paren{\frac{\p\alpha}{\p A}}_{M,Q}\ , \label{eq:int1}\\
\paren{\frac{\p\alpha p}{\p Q}}_{A,M}&=-\paren{\frac{\p\alpha\Phi}{\p A}}_{M,Q}\ , \label{eq:int2}\\
\paren{\frac{\p\alpha}{\p Q}}_{A,M}&=-\paren{\frac{\p\alpha\Phi}{\p M}}_{A,Q}\ . \label{eq:int3}
\end{align}
As it will be seen next, the three equations of state required to describe the system can be almost completely determined from the integrability conditions.  
In what follows, it is convenient to use the horizon radii as variables, in terms of which the line element \eqref{eq:dsRN} can be written as
\be
\dd s^2=-g_{00}\paren{r_+,r_-,r}\dd t^2+\frac{\dd r^2}{g_{00}\paren{r_+,r_-,r}}+r^2\dd\Omega^2\ ,
\en
with
\be
g_{00}\paren{r_+,r_-,r}=\paren{1-\frac{r_+}{r}}\paren{1-\frac{r_-}{r}}\ ,
\en
and
\be \label{eq:mQ}
m=\frac{1}{2}\paren{r_++r_-}\quad ; \quad Q^2=r_+r_-\ .
\en
The redshift function of the shell, given by 
\be
\label{eq:redshift}
k=\sqrt{g_{00}(a)}=\sqrt{\paren{1-\frac{r_+}{a}}\paren{1-\frac{r_-}{a}}}\ ,
\en
can be used to rewrite the material mass and pressure as follows:
\begin{align}
\label{Mk}
M^2&=4a^2k^2\ , \\
\label{pk}
p&=\frac{a^2\paren{1+k^2}-r_+r_-}{8\pi a^3k}\ .
\end{align}
Furthermore, let us note that
\be \label{eq:idp}
\paren{\frac{\p M}{\p a}}_{r_+,r_-}=-8\pi a p\ .
\en
Now, returning to Eq. \eqref{eq:int1} and considering that $A=4\pi a^2$, we have
\begin{align}
\frac{1}{8\pi a}\paren{\frac{\p\alpha}{\p a}}_{M,Q}&=\paren{\frac{\p \alpha p}{\p M}}_{a,Q} \label{eq:int1.1}\\
&=\alpha\paren{\frac{\p p}{\p M}}_{a,Q}+p\paren{\frac{\p\alpha}{\p M}}_{a,Q}, \label{eq:int1.2}\\
\frac{1}{8\pi a}\left[\paren{\frac{\p\alpha}{\p a}}_{r_+,r_-}-\paren{\frac{\p\alpha}{\p M}}_{Q,a}\paren{\frac{\p M}{\p a}}_{r_+,r_-}\right] &=\alpha\paren{\frac{\p p}{\p M}}_{a,Q}+p\paren{\frac{\p\alpha}{\p M}}_{a,Q}, \label{eq:int1.3}\\
\frac{1}{8\pi a}\paren{\frac{\p\alpha}{\p a}}_{r_+,r_-}&=\alpha\paren{\frac{\p p}{\p M}}_{Q,a}, \label{eq:int1.4}
\end{align}
where from \eqref{eq:int1.2} to \eqref{eq:int1.3} we have used the identity \eqref{eq:idalpha} and from \eqref{eq:int1.3} to \eqref{eq:int1.4} we have used \eqref{eq:idp}. Finally, the integrability condition \eqref{eq:int1} reads
\be
\paren{\frac{\p\alpha}{\p a}}_{r_+,r_-}=\alpha\frac{a^2\paren{1-k^2}-r_+r_-}{2a^3k^2}\ ,
\en
which has the general solution 
\be
\alpha=\mathrm{a}\paren{r_+,r_-}k\ ,
\en
where $\mathrm{a}(r_+,r_-)\equiv\alpha\paren{r_+,r_-,\infty}$ is an arbitrary function that represents the inverse temperature of the matter on a shell with infinite radius, and $k$ is the redshift factor defined in Eq. \eqref{eq:redshift}. Remarkably, this expression coincides with the one obtained in \cite{Lemos2015} for the temperature of a charged thin shell .

To obtain the equation of state corresponding to the electric potential $\Phi$, let us start by noting that 
\be
\paren{\frac{\p M}{\p A}}_{r_+,r_-}=-p\ .
\en
From the integrability conditions \eqref{eq:int1}-\eqref{eq:int3}, the  differential equation
\be
\paren{\frac{\p p}{\p Q}}_{A,M}+\paren{\frac{\p \Phi}{\p A}}_{M,Q}+\paren{\frac{\p M}{\p A}}_{r_+,r_-}\paren{\frac{\p\Phi}{\p M}}_{A,Q}+\Phi\paren{\frac{\p p}{\p M}}_{A,Q}=0\ 
\en
follows. 
Using \eqref{eq:idPhi}, this equation can be written as
\be
\paren{\frac{\p p}{\p Q}}_{a,M}+\frac{1}{8\pi a}\paren{\frac{\p\Phi}{\p a}}_{r_+,r_-}+\Phi\paren{\frac{\p p}{\p M}}_{a,Q}=0\ .
\en
Upon substitution of the corresponding derivatives,  the following differential equation is obtained:
\be \label{eq:intPhi}
a^2\paren{\frac{\p\Phi k}{\p a}}_{r_+,r_-}-2\sqrt{r_+r_-}=0\ ,
\en
which has the solution
\be\label{eq:solPhi}
\Phi(r_+,r_-,a)=\frac{\phi\paren{r_+,r_-}-\frac{2\sqrt{r_+r_-}}{a}}{k}\ ,
\en
where the function $\phi\paren{r_+,r_-}\equiv\Phi\paren{r_+,r_-,\infty}$ is the electric potential for $a=\infty$. It is expedient to define the function $\mathrm{c}\paren{r_+,r_-}\equiv\phi\paren{r_+,r_-}/\sqrt{r_+r_-}$ such that the electric potential takes the form
\be
\Phi\paren{r_+,r_-,a}=\frac{\sqrt{r_+r_-}}{k}\brac{\mathrm{c}\brac{r_+,r_-}-\frac{2}{a}}\ .
\en
The electric potential $\Phi$ can be thought as the difference between the electric potential of the shell if the radius were infinite and the electric potential of a shell of radius $a$, blueshifted by $k$. However, note that unlike the case of a spherical thin shell joining Minkowski and RN spacetime \cite{Lemos2015}, in our study we have two RN regions glued together, a fact that explains the factor of 2 in the last term of the Eq. \eqref{eq:solPhi}.

With the expressions of the temperature and electrostatic potential at our disposal, we can readily write $dS$ as (see Appendix \ref{app:dS} for details)
\be \label{eq:firstlaw2}
d S=\mathrm{a}\paren{r_+,r_-}\brac{1-\frac{\mathrm{c}\paren{r_+,r_-}r_-}{2}} d r_++\mathrm{a}\paren{r_+,r_-}\brac{1-\frac{\mathrm{c}\paren{r_+,r_-}r_+}{2}} d r_-\ ,
\en
which is a function only of $r_+$ and $r_-$. In order for $dS$ to be an exact differential,  the following integrability conditions should be satisfied:
\be \label{eq:intcondshell}
\frac{\p\mathrm{a}}{\p r_-}\paren{1-\frac{\mathrm{c}r_-}{2}}-\frac{\p\mathrm{a}}{\p r_+}\paren{1-\frac{\mathrm{c}r_+}{2}}=\frac{\mathrm{a}}{2}\frac{\p \mathrm{c}}{\p r_-}r_--\frac{\mathrm{a}}{2}\frac{\p \mathrm{c}}{\p r_+}r_+\ ,
\en
which coincides with the corresponding equation for the charged thin-shell, upon the rescaling $\mathrm{c}\to \mathrm{c}/2$ in the electrostatic potential \cite{Lemos2015}. 
By specifying either the inverse temperature or the electrostatic potential, the solution of this equation is used to obtain the entropy  from Eq. \eqref{eq:firstlaw2}.

\subsection{Thermodynamics of the TSW for a Hawking-type entropy}

Let us assume that the inverse temperature is given by
\be \label{eq:invtemp}
\mathrm{a}=\gamma\frac{r_+^2}{r_+-r_-}\ ,
\en
with $\gamma$ being a  parameter associated to the matter content of the shell. Substituting Eq. \eqref{eq:invtemp} into Eq. \eqref{eq:intcondshell} still results in some arbitrariness for the function $\mathrm{c}\paren{r_+,r_-}$. The \textit{Ansatz}
$\mathrm{c}=b/r_+$ solves the integrability condition for $b=2$, yielding
\be \label{eq:cfunct}
\mathrm{c}=\frac{2}{r_+}.
\en
With the choices of the functions $\mathrm{a}\paren{r_+,r_-}$ and $\mathrm{c}\paren{r_+,r_-}$ in Eqs. \eqref{eq:invtemp} and \eqref{eq:cfunct}, 
the entropy differential in Eq. \eqref{eq:firstlaw2} takes the form 
\be \label{eq:dSrp}
dS=\gamma r_+dr_+\ ,
\en
which can be directly integrated to obtain $S=S_0+\gamma r_+^2/2$, where $S_0$ is an integration constant, which we set to zero in order to have a null entropy when $m$ and $Q$ (hence $r_+$) go to zero. It follows that
\be \label{eq:entropy}
S=\frac{\gamma r_+^2}{2}\ ,
\en
where $r_+ = r_+(a_0, M, Q)$. As discussed in \cite{Lemos2015}, the constant $\gamma$ can be determined independently of the matter content in the limit $a_0\rightarrow r_+$, in which the temperature of the shell should coincide with the Hawking temperature, thus yielding $\gamma = 4\pi$. For values of $a_0$ other than $r_+$,
the value of $\gamma$ is not known a priori, the only requirement is that $\gamma > 0$ in order to have a positive entropy.

We can stress that the choice of the temperature function in Eq. \eqref{eq:invtemp} is only meaningful for the case where $m\ge Q$, in which the horizon $r_+$ exists and the radius of the throat  obeys $a_0>r_+$. Therefore, Eq. \eqref{eq:invtemp} is not suited for systems with $m<Q$. Taking into account that the dynamical stability region of TSWs governed by a linear equation of state is larger in the $m<Q$ regime (see Fig.\ref{fig:dyn_stab}), it is convenient to study a more general form for the entropy so that this region can be adequately taken into account.

\subsection{Thermodynamics  of the TSW for a power law entropy }

Let us next consider a more general case in which the inverse temperature depends on the ADM mass and obeys a power law equation of state. Recalling that $2m=r_++r_-$, the following \textit{Ansatz} for the inverse temperature will be adopted \cite{Lemos2015} 
\be \label{eq:invtempgen}
\mathrm{a}\paren{r_+,r_-}=\nu\ \paren{r_++r_-}^{\delta}\ ,
\en
where $\nu$ and $\delta$ are parameters related to the matter content of the shell. Using Eq. \eqref{eq:invtempgen} in the integrability condition \eqref{eq:intcondshell}, it follows that 
\be
\frac{\delta\paren{r_+-r_-} }{\paren{r_++r_-}^\delta}\mathrm{c}\paren{r_+,r_-}=\frac{\p \mathrm{c}\paren{r_+,r_-}}{\p r_-}r_{-}-\frac{\p \mathrm{c}\paren{r_+,r_-}}{\p r_+}r_+\ .
\en
The general solution  of this equation has the form $\mathrm{c}=f\paren{r_+r_-}/\paren{r_++r_-}^\delta$ where $f\paren{r_+r_-}$ is an arbitrary function of the product $r_+r_-$ that is ultimately related to the electric charge. We will  choose a power law for the electrostatic potential, such that
\be \label{eq:genc}
\mathrm{c}\paren{r_+,r_-}=\xi\frac{\paren{r_+r_-}^{\mu}}{\paren{r_++r_-}^\delta}\ ,
\en
where $\xi$ and $\mu$ are parameters  related to the matter content of the shell. 
With the replacement of Eqs. \eqref{eq:invtempgen} and \eqref{eq:genc} into Eq. \eqref{eq:firstlaw2}, after integration, we obtain the entropy
\be \label{eq:entropygen1}
S = S_0 + \nu \left[ \frac{\paren{r_+ + r_-}^{\delta +1}}{\paren{\delta +1}} - \frac{ \xi }{2}\frac{\paren{r_+r_-}^{\mu +1}}{\paren{\mu +1}} \right] \ ,
\en
where $S_0$ is an integration constant. Using Eq. \eqref{eq:mQ}, we arrive to
\be \label{eq:entropygen2}
S = S_0 + \nu \left[ \frac{\paren{2m}^{\delta +1}}{\paren{\delta +1}} - \frac{ \xi }{2}\frac{\paren{Q^2}^{\mu +1}}{\paren{\mu +1}} \right] \ ,
\en
in such a way that the definition of the entropy can be extended to the overcharged (\textit{i.e.} $m<Q$) case.
Let us next place some physically relevant restrictions on the parameters appearing in the expression for the entropy.
As in the previous section, the integration constant $S_0$ is set to zero, so that in the absence of horizons $(m=Q=0)$ the entropy is null. 
The adoption of $\nu >0$ guarantees a positive entropy when $Q \to 0$ (for fixed $m$). 
The parameters $\delta$ and $\mu$ have a lower bound given by $\delta>-1$ and $\mu>-1$, regardless of the value of $\nu$, to ensure that $S$ does not diverge in the limits $m\to0$ and $Q\to0$.
The parameter $\xi$ in Eq. \eqref{eq:entropygen1} sets the overall sign of the thermodynamical electric potential $\mathrm{c}\paren{r_+,r_-}$ in Eq. \eqref{eq:genc}. We set $\xi>0$ in order to have a positive potential, as in Ref. \cite{Lemos2015}.

\section{Conditions for thermodynamical and complete stability}
\label{condsstab}

Next, let us present the necessary steps to obtain the constraints imposed by the thermodynamical stability on the TSW. We will follow closely the presentation in \cite{Lemos2015}, adapted to the charged case as in \cite{Reyes2022}.

The regions of thermodynamical stability are determined by the following conditions \cite{Lemos2015,Callen1985}:
\be\label{eq:cond1}
\paren{\frac{\p^2S}{\p M^2}}_{A,Q}\le0\ ,
\en
\be\label{eq:cond2}
\paren{\frac{\p^2S}{\p A^2}}_{M,Q}\le0\ ,
\en
\be\label{eq:cond3}
\paren{\frac{\p^2S}{\p Q^2}}_{M,A}\le0\ ,
\en
\be\label{eq:cond4}
\paren{\frac{\p^2S}{\p M^2}}\paren{\frac{\p^2S}{\p A^2}}-\paren{\frac{\p^2S}{\p M\p A}}^2\ge0\ ,
\en
\be\label{eq:cond5}
\paren{\frac{\p^2S}{\p A^2}}\paren{\frac{\p^2 S}{\p Q^2}}-\paren{\frac{\p^2S}{\p A\p Q}}^2\ge0\ ,
\en
\be\label{eq:cond6}
\paren{\frac{\p^2S}{\p M^2}}\paren{\frac{\p^2S}{\p Q^2}}-\paren{\frac{\p^2S}{\p M\p Q}}^2\ge0\ ,
\en
and
\be\label{eq:cond7}
\paren{\frac{\p^2S}{\p M^2}}\paren{\frac{\p^2S}{\p Q\p A}}-\paren{\frac{\p^2S}{\p M\p A}}\paren{\frac{\p^2 S}{\p M\p Q}}\ge0\ ,
\en
which are obtained by ensuring that the variations on mass, area, and charge on the entropy of the two subsystems do not lead to inhomogeneities and a consequent phase transition.

Taking into account that the charge of the system depends on the matter content via the relation given in Eq. \eqref{eq:beta}, the thermodynamical stability conditions reduce to the set (see Appendix \ref{derivcondsstab}): \begin{subequations} \label{eq:thermstab}
\be
S_{AA} \le0\ , \label{eq:thermstab2}
\en
\be
\beta^2S_{QQ}+2\beta S_{MQ}+S_{MM} \le0\ , \label{eq:thermstab1}
\en
and
\be
S_{AA}\paren{S_{MM}+\beta^2S_{QQ}+2\beta S_{MQ}}-\paren{\beta S_{AQ}+S_{AM}}^2 \ge0\ ,\label{eq:thermstab3}
\en
\end{subequations}
where the subscript ${X}$ means derivative with respect to $X$. These conditions, when  applied to a given entropy function, determine the regions of thermodynamical stability for the system under consideration, in terms of $a_0$, $\beta$, and the parameter $\kappa$ of the linear EOS for the pressure.

\subsection{Thermodynamical stability for the TSW for a Hawking type entropy }

Before the examination of the charged case, let us present a simpler system, which is obtained from the charged TSW by taking the $Q=0$ limit, leading to the gluing of two Schwarzschild spacetimes, with $r_+ = r_g = 2m>0$ and a throat radius $a_0>2m$. Setting $\beta=0$ in Eqs. \eqref{eq:thermstab}, they yield
\begin{subequations} \label{eq:redthermstab}
\be
S_{AA}\le 0\ , \label{eq:redthermstab2}
\en
\be
S_{MM}\le 0\ , \label{eq:redthermstab1}
\en
and
\be
S_{AA}S_{MM}-S_{AM}^2\ge0\ . \label{eq:redthermstab3}
\en
\end{subequations}
We also assume that the entropy is given by Eq. \eqref{eq:entropy}. The stability condition \eqref{eq:redthermstab1} reads 
\be
\frac{\gamma(3\pi M^2-A)}{2A} \le 0 \ ,
\label{eq:redthermstab1H1}
\en
which, with the help of Eq. \eqref{eq:mADM}, implies that $a_0\le 3m$, just as in the case of the uncharged thin shell \cite{Lemos2015,Martinez1996}. Using relations \eqref{eq:ax} and \eqref{eq:Mx}, the inequality \eqref{eq:redthermstab1H1} takes the form
\be
\frac{2 \gamma (\kappa +1)}{-(4 \kappa +1)} \le 0 \ ,
\label{eq:redthermstab1H2}
\en
which, since $-(4\kappa +1)>0$ by Eq. \eqref{eq:physcond1}, can be restated as $\kappa \le -1$. The combined stability condition \eqref{eq:redthermstab3} yields
\be
\frac{\gamma ^2\pi M^4(\pi M^2-A)}{8A^4} \ge 0 \ ,
\en
or equivalently
\be
\frac{\gamma ^2 \sigma_0 ^2 (2 \kappa +1)}{4 (4 \kappa +1)^2} \ge 0 \ ,
\en
which always requires that $k^2\ge 1$, but it can not be fulfilled, because $a_0>2m>0$ forces that $k=\sqrt{1-2m/a_0}<1$. Finally, condition \eqref{eq:redthermstab2} implies that the inequality 
\be
\frac{\gamma}{4A^3}M^4\pi \le0\ ,
\label{eq:redthermstab2beta0}
\en
or rewritten in the form
\be
\frac{\gamma  \sigma_0 ^2}{-4(4\kappa +1)} \le0\ ,
\label{eq:redthermstab2beta0rew}
\en
is never satisfied, due to the strict positivity of all the parameters involved. Hence, all the {uncharged} configurations are thermodynamically unstable. 

Let us now analyze the set of stability conditions \eqref{eq:thermstab} for the charged case in which $\beta \neq 0$ and the entropy is given by Eq. \eqref{eq:entropy}. The inequalities in this case respectively take the form
\begin{eqnarray}
&& \frac{\gamma  \sigma _0 ^2}{8 \left(\psi-4 \kappa \right) \omega^{3}} \left[8 \kappa ^3 \left(16 \beta ^4+8 \beta ^2-1\right) +4 \psi ^2 \kappa ^2 \left(\omega-3\right)
\right. \nonumber \\
&& \left. +2 \psi ^2 \kappa \left(6 \beta ^2+2 \omega-3\right)-\psi ^3 \left(2 \beta ^2+\omega-1\right)\right] \le 0 \ ,
\end{eqnarray}
\begin{eqnarray}
&& \frac{\gamma}{\left(\psi -4 \kappa\right) \omega^{3}}  \left\{ -8 \kappa^4+4 \kappa^3 \left(24 \beta ^4+14 \beta ^2+\omega-5\right) +\kappa^2 \left[ 48 \beta ^4 \left(\omega-2\right)-4 \beta ^2 \left(7 \omega-24\right) +8 \omega-18\right] \right. \nonumber \\
&& \left. +\psi \kappa \left[ 16 \beta^4+4 \beta ^2 \left(3 \omega-8\right)-5 \omega+7\right] -\psi ^2 \left[ 4 \beta ^4+\beta ^2 \left(3 \omega-5\right)-\omega+1\right] \right\} \le 0 \ ,
\end{eqnarray}
and
\begin{eqnarray}
&& \frac{\gamma ^2 \sigma _0 ^2}{8 \left(\psi -4 \kappa \right)^2 \omega^4} 
\left\{-32 \kappa^5 \left(8 \beta ^2-1\right)+16 \kappa^4 \left[ 128 \beta ^8+96 \beta ^4+\beta ^2 \left(8 \omega-46\right)-\omega+5\right] \right. \nonumber \\
&& +16 \psi \kappa^3 \left[ 8\beta ^6 \left(2 \omega-1\right)+\beta ^4 \left(4 \omega-62\right) +\beta ^2 \left(36-11 \omega\right)+2 \omega-5\right] \nonumber \\
&& -8 \psi ^2 \kappa ^2 \left[ 4 \beta ^6+\beta ^4 \left(6 \omega-41\right)+6 \beta ^2 \left(5-2 \omega \right)+3 \omega -5\right] -2 \psi ^3 \kappa \left[ 8 \beta ^6-2 \beta ^4 \left(\omega-15\right) \right. \nonumber \\
&&  \left. \left. +14 \beta ^2 \left(\omega-2\right) -4 \omega+5\right] -\psi ^4 \left[ 2 \beta ^4 \left(\omega -4\right) +\beta ^2 \left(6-4 \omega\right)+\omega-1\right] \right\} \ge 0 \ ,
\end{eqnarray}
where $\psi \equiv 4 \beta ^2 -1$ and $\omega \equiv \sqrt{(2 \kappa+1)^2-4 \beta ^2}$. As $\gamma >0$, $\sigma_0^2>0$, and, from Eq. \eqref{eq:physcond1}, $\psi - 4 \kappa >0$, the left overall factor in each of these equations is positive and the fulfillment of the inequalities is completely determined by the second factor, which is independent of $a_0$. However, the corresponding expressions are cumbersome, so we present the results graphically in Fig. \ref{fig:th_stab}, which displays the region in the plane $(\beta,\kappa)$ where all the three inequalities are simultaneously satisfied. This figure also shows the stability conditions discussed above for the uncharged case ($\beta=0$). Unlike the latter case, thermodynamically stable configurations are possible for non-zero charge. In Fig. \ref{fig:th_stab} the dynamical stability region is also shown for comparison. It follows from the plot that, for a linear equation of state  and a Hawking type entropy, there are no configurations both dynamically and thermodynamically stable. 

Next, we will explore the consequences of adopting the more general expression for the entropy given in Eq. \eqref{eq:entropygen2}, which allows the existence of thermodynamically stable regions within the overcharged domain of the configurations.

\begin{figure}[t!]
\centering
\includegraphics[width=0.4\textwidth]{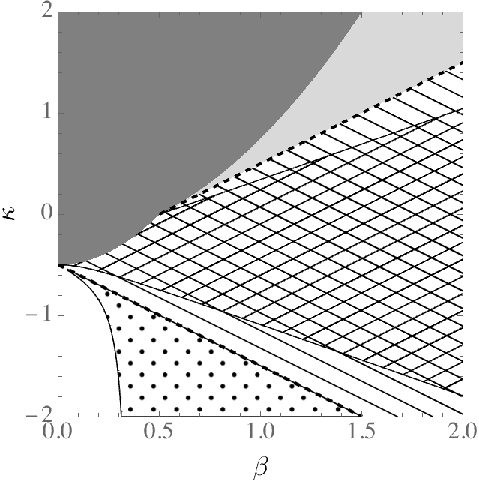}
\caption{Regions of thermodynamical stability in the $\paren{\beta,x}$ plane for the Hawking type entropy, with arbitrary $\gamma>0$. The thermodynamical stability corresponds to the dot meshed region and the dynamical stability to the upward-line meshed one. The downward-line meshed region shows the overcharged case, bounded by the dashed lines $ \kappa = \pm \beta -1/2$ for which  $Q=m$. Configurations in the light gray zone have $a_0<r_+$ and are not considered in this work.  The medium gray region is non-physical (see text).}
\label{fig:th_stab}
\end{figure}

\subsection{{Thermodynamical stability for the power law entropy}}

The entropy displayed in Eq. \eqref{eq:entropygen2} has more parametric freedom than the one explored in the previous subsection. Let us first study the uncharged case, by taking the Q = 0 limit to obtain the TSW with two Schwarzschild submanifolds, and a throat radius $a_0 > r_+ =2m>0$. Setting $\beta = 0$ in Eq. \eqref{eq:entropygen2} and using Eqs. \eqref{eq:redthermstab2}, \eqref{eq:redthermstab1}, and \eqref{eq:redthermstab3}, we obtain 
\be
\frac{ \pi ^{1-\delta }}{4} \nu \left| \sigma_0 \right| ^{3-\delta } (-4 \kappa-1)^{(1-3 \delta )/2} (-2 \kappa-1)^{\delta -1} \left[4 (\delta -1) \kappa^2+1\right] \le 0 \ ,
\en
\be
\pi ^{1-\delta } \nu \left| \sigma_0 \right| ^{1-\delta }  (-4 \kappa-1)^{(1-3 \delta )/2} (-2 \kappa-1)^{\delta -1} (\delta +2 \kappa+1) \le 0 \ ,
\en
and
\be
\frac{ \pi ^{2\left(1-\delta \right)} }{4} \nu ^2 \left| \sigma_0 \right| ^{4-2 \delta } (-4 \kappa-1)^{1-3 \delta } (-2 \kappa-1)^{2 \delta -1} \left[ 2 (1-\delta) \kappa (2 \kappa +1) -\delta \right]  \ge 0.
\en
The bounds on $\kappa$ from each of the stability conditions are given by
\be
\left| \kappa \right| \ge\frac{1}{2\sqrt{1-\delta}}\ \text{for} \ \delta <1\ , \label{eq:firstboundkappa1}
\en
\be
\kappa\le-\frac{\delta+1}{2}\ ,
\en
and
\be
\begin{cases}
\kappa\in\mathbb{R}\ \text{for} \ \delta \le -1/3 \\
\left| \kappa +\frac{1}{4} \right| \ge \frac{1}{4}\sqrt{\frac{1+3\delta}{1-\delta }}\ \text{for} \ -1/3 <\delta <1  \ , \label{eq:thirdboundkappa}
\end{cases}
\en
respectively. The above conditions should be matched with $\delta>-1$ that comes from the entropy in Eq. \eqref{eq:entropygen2} as well as the physical requirement that $\kappa \leq -1/2$ resulting from Eq. \eqref{eq:physcond2}. 
Taking these constraints into account and putting all together, the possible values of $\kappa$ that  simultaneously satisfy the three stability conditions read
\be
\begin{cases}
\kappa \le -1/2  \ \text{for} \  -1<\delta < 0 \\
\kappa \le -\frac{1}{4}\paren{1+\sqrt{\frac{1+3\delta}{1-\delta }}}\  \text{for} \  0 \le \delta <1 \ .
\end{cases}
\en

In the presence of charge, \textit{i.e.} $\beta \neq 0$, it is convenient to start by constraining the physically permitted domains for the parameters. We first consider the thermodynamical stability condition given by Eq. \eqref{eq:thermstab2}, which does not involve the parameters $\xi$ and $\mu$. Given the expressions for the mass and throat radius in Eqs. \eqref{eq:ax}-\eqref{eq:mx}, the inequality \eqref{eq:thermstab2} takes the form
\be \label{eq:SAAineq}
\frac{\pi^{1-\delta} \nu |\sigma_0|^{3-\delta} \left[ 4\paren{\beta^2-\kappa}-1\right] ^{(1-3\delta )/2}}{4\paren{4\beta^2-2\kappa-1}^{1-\delta }} \left[\paren{4\beta^2-1}^2+4\kappa^2\paren{\delta-1}\right] \le0\ .
\en
From the physical requirements for the parameters in Eqs. \eqref{eq:physcond1} and \eqref{eq:physcond2}, and considering that $\nu>0$, the sign of the left hand side of this expression is determined by the sign of the last factor between square brackets. From this factor it follows that 
\be\label{eq:deltanup}
\delta \leq 1-\frac{1}{4\kappa^2}(4\beta^2-1) ^2,
\en
for any $\kappa \neq 0$. In particular, note that the bound in Eq. \eqref{eq:deltanup} implies that $\delta\le1$, with the equality holding only over the line $\beta=1/2$, \textit{i.e.} for $\delta=1$ the fulfillment is confined to this line. Taking into account that $\delta>-1$ from the definition of the entropy in \eqref{eq:entropygen2}, there is a lower and an upper limit for the exponent $\delta$, given by $-1<\delta \le 1$, which is a necessary condition for stability. The remaining thermodynamical stability conditions, namely Eqs. \eqref{eq:thermstab1} and \eqref{eq:thermstab3}, involve derivatives with respect to $M$ and $Q$, and hence the parameters $\xi$ and $\mu$ will appear explicitly in the corresponding expressions. Choosing $\delta=2\mu+1$ in order for $\xi$ to be a dimensionless constant, the conditions \eqref{eq:thermstab1} and \eqref{eq:thermstab3} read 
\be
\pi ^{1-\delta } \nu |\sigma_0|^{1-\delta} \left[ 4\left( \beta ^2- \kappa \right)-1\right]^{1-\delta } \left[ \frac{\left(4 \beta ^2-1\right) \left[\left(4 \beta ^2-1\right) (1+\delta )-2 \kappa\right]}{\left(4 \beta ^2-2 \kappa-1\right)^{1-\delta }\left[ 4 \right( \beta ^2- \kappa \left) -1\right] ^{(1+\delta )/2}} - \delta \xi\beta^{1+\delta } \right] \le 0
\en
and
\begin{eqnarray}
&& \frac{\pi ^{2 (1-\delta )} \nu ^2 \left| \sigma_0 \right| ^{4-2\delta } \left[4 \left(\beta ^2- \kappa \right) -1\right]^{1-3 \delta }}{4 \left(4 \beta ^2-2 \kappa -1\right)^{1-3 \delta }}
\left\{ - \delta  \xi \beta^{1+\delta } \left[ 4 \left(\beta ^2-\kappa \right)-1\right]^{(1+\delta )/2} \left[ \left(4 \beta ^2-1\right)^2+4 \kappa^2 (\delta -1)\right] \right. \nonumber \\
&& \left. +\left(4 \beta ^2-1\right) \left(4 \beta ^2-2 \kappa-1\right)^{\delta }  
 \left[\left(4 \beta ^2-1\right)^2 \delta +2\kappa (1-\delta ) \left(4 \beta ^2-2 \kappa-1\right)\right]  \right\} \ge 0  \ ,
\end{eqnarray}
respectively. Note that these conditions are independent of $a_0$, since the throat radius only appears in them through $|\sigma_0|>0$. For $\delta =1$ the thermodynamical stability is restricted to the line $\beta =1/2$, where all conditions are satisfied; in this case, by using Eqs. \eqref{eq:physcond1} and \eqref{eq:physcond2} we require that $\kappa <0$, and also that $\kappa \le -\xi/4$ to avoid a negative entropy. When $-1 < \delta <1$, we can numerically study the possible configurations in the parameter space that fulfill all the three inequalities, subject to the constrains given by Eqs. \eqref{eq:physcond1} and \eqref{eq:physcond2}. 

\begin{figure}[t!]
\centering
\includegraphics[width=0.8\textwidth]{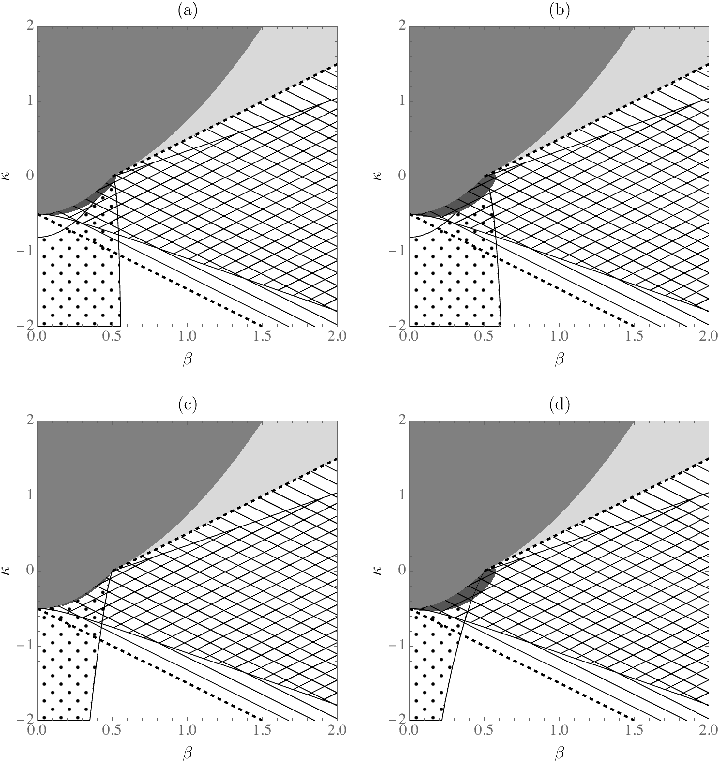}
\caption{Regions of thermodynamical stability in the $\paren{\beta,\kappa}$ plane for the power law entropy. In all plots, $\nu > 0$ is arbitrary, while in (a) $\delta=1/2$, $\mu=-1/4$,  and $\xi = 1/2$; (b)  $\delta=1/2$, $\mu=-1/4$, and $\xi = 1$; (c)  $\delta=-1/2$, $\mu=-3/4$, and  $\xi = 1/2$; (d)  $\delta=-1/2$, $\mu=-3/4$, and $\xi = 1$. The thermodynamically stable region is represented by a dot mesh while the dynamically stable region by a upward-line mesh. The downward-line meshed region corresponds to the overcharged case $Q > m$, bounded by the dashed lines $ \kappa = \pm \beta -1/2$ for which  $Q=m$. The light gray region is not allowed since $a < r_+ $ in there. The medium gray region is non-physical (see text). A small negative entropy zone appears, shown in dark gray, which is also considered non-physical.}
\label{fig:pstabentropygen}
\end{figure}

In order to present some examples, we adopt an arbitrary $\nu> 0$ and the relation $\delta = 2\mu + 1$ so the units of both terms of the entropy are the same and the parameter $\xi >0$ is dimensionless. Fig. \ref{fig:pstabentropygen} shows that termodinamically stable regions (where the three inequalities are valid) in the  $\paren{\beta,\kappa}$ plane are possible for suitable values of the parameters. The regions in this plane that do not satisfy Eqs. \eqref{eq:physcond1} and \eqref{eq:physcond2} are non-physical and should be excluded. Our construction also takes out the region for which $a \le r_+$. In all plots of Fig. \ref{fig:pstabentropygen} we have taken arbitrary $\nu > 0$, while in (a) $\delta=1/2$, $\mu=-1/4$,  and $\xi = 1/2$; (b) $\delta=1/2$, $\mu=-1/4$, and $\xi = 1$; (c) $\delta=-1/2$, $\mu=-3/4$, and  $\xi = 1/2$; (d)  $\delta=-1/2$, $\mu=-3/4$, and $\xi = 1$. The dynamical stability is displayed again in Fig. \ref{fig:pstabentropygen}. We can observe a partial overlapping of the dynamically and thermodynamically stable regions for the overcharged case. The thermodynamical stability behaviour for other values of $\delta$ within the range $-1<\delta <0$ is qualitatively similar to $\delta = -1/2$ and for $0\le \delta <1$ to $\delta = 1/2$ . 

The plots in Fig. \ref{fig:pstabentropygen} show that, in spite of the fact that the intersection of the regions associated with the dynamical and thermodynamical stability is highly dependent on the values of the parameters chosen for the entropy,  it is always present for $-1<\delta<1$ as a small zone in the $(\beta,\kappa)$ plane.
It is worth noticing that all the completely stable configurations correspond to the overcharged case since, as shown in Sect. \ref{dinstab}, these are the only states that are dynamically stable for a linear equation of state.

\section{Conclusions} \label{sec:conclusions}

In the present work, the conditions for dynamical and thermodynamical stability for a charged TSW built by gluing two Reissner-Nordstrom geometries have been obtained. Regarding the dynamical stability for perturbations preserving the spherical symmetry, the adopted formalism allows to present the stability condition as
a simple inequality involving the ratio $\beta =Q/|M|$ between the charge $Q$ and the material mass $M$, and the relation arising from the barotropic equation of state $p=x(\sigma )\sigma $ between the pressure and the energy density. In particular, for a linear EOS, the wormhole is entirely characterized by the throat radius $a_0$, $\beta $, and constant $x(\sigma ) = \kappa $. In this case, we have found that the dynamical stability is independent of $a_0$ and the region of the parameter space $(\beta , \kappa )$ corresponding to stable configurations falls within the overcharged zone and widens for larger values of $\beta $.

Following \cite{Lemos2015}, we have developed a thermodynamical description for the matter on the TSW.\footnote{Let us emphasize that, despite representing a completely different physical situation, the thermodynamical description follows closely that of the charged thin-shell \cite{Lemos2015}.} In such a description, the expression for the entropy is obtained by a suitable choice of the inverse temperature (which also constraints the form of the electric potential) as a function of the extensive variables. Possible values for the parameters in the entropy were determined by imposing physically reasonable conditions. In order to study the corresponding stability, two convenient forms of the inverse temperature have been used. 

We have shown that the imposition of the thermodynamical stability conditions (given by inequalities for different combinations of second order derivatives of the entropy) gives the set of stable configurations. The restriction resulting from these conditions can be more or less severe, depending on the system under study and  on the entropy function. We have adopted a linear EOS for the pressure, so the thermodynamical features of the wormholes are fully determined by $a_0$, $\beta $, and $\kappa $, but in both entropy models the stability is independent of $a_0$. In particular, for the Hawking type entropy that depends quadratically on $r_+$, we have shown that there are no thermodynamically stable states for the non-charged case (namely, a Schwarzschild-Schwarzschild TSW), and there are thermodynamically but not completely stable configurations for the charged case.

The consideration of a more general entropy with a power law function of the ADM mass and the charge leads to more interesting results. In this case, thermodynamically stable but dynamically unstable configurations without charge are possible. In the presence of charge, we have shown that the thermodynamically stable region in the parameter space $(\beta, \kappa )$ extends into the overcharged zone where dynamical stability is also possible, resulting in equilibrium configurations that are completely stable. This small region of complete stability appears for the exponent $\delta$ within the range  $-1<\delta<1$, negative $\kappa $ (\textit{i.e.} positive pressure), and low values of $\beta $.

Let us point out that the entropy functions used here are such that the temperature is always positive in order to have a positive entropy. It would be interesting to study cases that allow negative temperatures without the undesirable feature of a negative entropy, and their relation with the formalism developed in \cite{MartinMoruno2009} for (static or dynamic) wormholes with continuous distributions of matter (as opposed to a thin shell) as a source.

\begin{acknowledgments}
This work was supported by CONICET (EFE and GFA) and by National Council for Scientific and Technological Development - CNPq and FAPERJ - Fundaçāo Carlos Chagas Filho de Amparo à Pesquisa do Estado do Rio de Janeiro, Processo SEI 260003/014960/2023 (MLP and SEPB).
    
\strut
    
\textbf{Data Availability Statement}: This manuscript has no associated data or the data will not be deposited. [Authors’ comment: All data generated or analysed during this study are included in this published article.]   
\end{acknowledgments}

\appendix

\section{Deriving the thermodynamical stability conditions}
\label{derivcondsstab}

In this appendix, we review the derivation of the thermodynamical stability conditions \cite{Lemos2015,Callen1985,Reyes2022}. The local thermodynamical stability conditions ensure that a particular subsystem does not present inhomogeneities that might lead to a phase transition. The latter condition can be restated as imposing that the entropy is an extremum, \textit{i.e.}, $dS=0$, and a maximum, \textit{i.e.}, $d^2S<0$.

We begin by considering a system which can be divided into two subsystems, each with entropy $S$. In particular, let each of the subsystems be characterized by three independent variables: the internal energy $M$, its charge $Q$, and its area $A$. So $ S(M,A,Q)$ is the entropy of each of the subsystems. Now, let both systems exchange some amount of energy $\Delta M$, such that the resultant entropy of the whole is $S\paren{M+\Delta M,Q,A}+S\paren{M-\Delta M,Q,A}$. If the resultant entropy is greater than the initial entropy of both subsystems, then inhomogeneities tend to be formed and will eventually give rise to a phase transition. Therefore, the condition for thermodynamical stability upon the exchange of some energy $\Delta M$ can be stated as
\be \label{eq:ap1}
S(M+\Delta M,Q,A)+S(M-\Delta M,Q,A)\le 2S(M,Q,A)\ .
\en
Next, a Taylor expansion of Eq. \eqref{eq:ap1} up to second order in $\Delta M$, in the limit $\Delta M\to 0$ leads to the condition
\be \label{eq:apstab1}
 \left(\frac{\p^2S}{\p M^2}\right)_{A,Q}\le 0\ .
\en
Similarly, when considering exchanges of charge $\Delta Q$ and changes in the area $\Delta A$, separately results in
\begin{align}
\paren{\frac{\p^2S}{\p Q^2}}_{M,A}&\le0\ , \label{eq:apstab2}\\
\paren{\frac{\p^2S}{\p A^2}}_{M,Q}&\le0\ , \label{eq:apstab3}
\end{align}
respectively. 

Considering variations of two variables while taking one quantity fixed, it follows that 
\be \label{eq:apvarMQ1}
S_{MM}\paren{\Delta M}^2+S_{QQ}\paren{\Delta Q}^2+2S_{MQ}\Delta M\Delta Q\le0\ ,
\en
where we have adopted the notation $S_{XY}=\p^2 S/\p X\p Y$ for the case of constant $A$. We can multiply the latter expression by the relation \eqref{eq:apstab1}, obtaining
\be
\paren{S_{MM}\Delta M+S_{MQ}\Delta Q}^2+\Delta Q^2\paren{S_{MM}S_{QQ}-S_{MQ}^2}\ge0 \ .
\en
Noting that the first term is positive definite, we arrive to the inequality
\be \label{eq:apcon2a}
S_{MM}S_{QQ}-S_{MQ}^2\ge0\ ,
\en
establishing the thermodynamical stability condition for variations on the energy and charge. Next, considering variations on the energy and area, and performing the same steps as in the previous inequality, we obtain that the condition for thermodynamical stability of the system under changes in energy and area is
\be \label{eq:apcond2b}
S_{MM}S_{AA}-S_{MA}^2\ge0\ .
\en
Finally, one last possible variation of two parameters remains to be studied: the variation of area and charge. Following the same steps we obtain that the condition for thermodynamical stability reads
\be
S_{AA}S_{QQ}-S_{AQ}^2\ge0\ .
\en

There is another possible configuration to analyze: the variation of the three parameters comprising the system. By performing a Taylor expansion up to second order, we obtain that the entropy should obey the inequality
\be \label{eq:apfinalcondprev}
S_{MM}\paren{\Delta M}^2+S_{QQ}\paren{\Delta Q}^2+S_{AA}\paren{\Delta A}^2+2S_{MA}\Delta M\Delta A+2S_{MQ}\Delta M\Delta Q+2S_{QA}\Delta Q\Delta A\le0\ .
\en
Multiplying the latter by the condition \eqref{eq:apstab1} we can rearrange the resulting inequality as
\begin{align}
\paren{S_{MM}\Delta M+S_{MQ}\Delta Q+S_{MA} \Delta A}^2&+\paren{S_{MM}S_{QQ}-S_{MQ}^2}\paren{\Delta Q}^2+\paren{S_{MM}S_{AA}-S_{MA}^2}\paren{\Delta A}^2 \nonumber\\
&+2\paren{S_{MM}S_{QA}-S_{MQ}S_{MA}}\Delta A\Delta Q\ge 0 .
\end{align}
Note that in the latter the first term is positive definite and the two following terms in parenthesis are the conditions \eqref{eq:apcon2a} and \eqref{eq:apcond2b}. Thus, the thermodynamical stability condition for variations on the three parameters of the system reads.
\be \label{eq:apconfinal}
S_{MM}S_{QA}-S_{MQ}S_{MA}\ge0\ .
\en
Let us note, however, that the condition \eqref{eq:apconfinal} is not unique. In fact, alternatively using Eqs. \eqref{eq:apstab2} and \eqref{eq:apstab3} in Eq. \eqref{eq:apfinalcondprev} one arrives to
\begin{align}
    S_{QQ}S_{MA}-S_{MQ}S_{QA}\ge0\ , \label{eq:apconfinal2}\\
    S_{AA}S_{MQ}-S_{MA}S_{QA}\ge0\ . \label{eq:apconfinal3}
\end{align}
At first glance, it may seem that the conditions \eqref{eq:apconfinal}, \eqref{eq:apconfinal2} and \eqref{eq:apconfinal3} are independent despite all three being derived from the same initial condition \eqref{eq:apfinalcondprev}. In fact, it is possible to prove that using one of the latter conditions and the previous inequalities one arrives to one other. To illustrate this fact let us take the condition \eqref{eq:apconfinal} and multiply it by \eqref{eq:apstab3}, namely $S_{AA}\le0$. Therefore, we obtain
\be
S_{MM}S_{AA}S_{QA}-S_{MQ}S_{MA}S_{AA}\le0\ .
\en
We can further use the inequality \eqref{eq:apcond2b} in the latter and rearrange the terms in order to obtain the condition \eqref{eq:apconfinal3}. The same can be done with each of the final conditions \eqref{eq:apconfinal}-\eqref{eq:apconfinal3} and show that they are equivalent.

In the case in which the charge is a function of $M$, namely $Q=f(M)$, the system is characterized by $S(M,A,f(M))$.
Proceeding as before, the conditions for thermodynamical equilibrium are
\be
S_{AA}\le0\ ,
\en
\be
f''S_{f}+\paren{f'}^2S_{ff}+2f'S_{fM}+S_{MM}\le0\ ,
\en
and
\be
S_{AA}\brac{S_{MM}+\paren{f'}^2S_{ff}+2S_{Mf}f'+f''S_f}-\paren{f'S_{Af}+S_{MA}}^2\ge0\ .
\en
where $f'$ stands for $df/dM$.

\section{Entropy differential} \label{app:dS}

In this appendix, we start by reviewing the Maxwell relations, commonly used in the treatment of thermodynamic systems, adapted to the notation used in this paper (suitable to the treatment of gravitating systems). First, the differential of a function $\psi\paren{x,y,z}$ is
\be \label{eq:rel1}
d\psi=\paren{\frac{\p\psi}{\p x}}_{y,z}dx+\paren{\frac{\p\psi}{\p y}}_{x,z}dy+\paren{\frac{\p\psi}{\p z}}_{x,y}dz\ .
\en
In turn, considering that
\be\label{eq:rel2}
d\psi=\paren{\frac{\p\psi}{\p u}}_{v,w}du+\paren{\frac{\p\psi}{\p v}}_{u,w}dv+\paren{\frac{\p\psi}{\p w}}_{u,v}dw
\en
with
\be\label{eq:rel3}
\paren{\frac{\p\psi}{\p u}}_{v,w}=\paren{\frac{\p\psi}{\p x}}_{y,z}\paren{\frac{\p x}{\p u}}_{v,w}+\paren{\frac{\p\psi}{\p y}}_{x,z}\paren{\frac{\p y}{\p u}}_{v,w}+\paren{\frac{\p\psi}{\p z}}_{x,y}\paren{\frac{\p z}{\p u}}_{v,w}
\en
and so forth. Choosing $\psi=\Phi$, $x=M$, $y=Q$, $z=R$ and $u=r_+$, $v=r_-$ and $w=R$ in the above relations we find
\be\label{eq:idPhi}
\paren{\frac{\p\Phi}{\p R}}_{M,Q}=\paren{\frac{\p\Phi}{\p R}}_{r_+,r_-}-\paren{\frac{\p\Phi}{\p M}}_{Q,R}\paren{\frac{\p M}{\p R}}_{r_+,r_-}\ .
\en
Now, taking $\psi=\alpha$, $\paren{x,y,z}=\paren{M,Q,R}$ and $\paren{u,v,w}=\paren{r_+,r_-,R}$ we obtain
\be\label{eq:idalpha}
\paren{\frac{\p\alpha}{\p R}}_{r_+,r_-}=\paren{\frac{\p\alpha}{\p M}}_{Q,R}\paren{\frac{\p M}{\p R}}_{r_+,r_-}+\paren{\frac{\p\alpha}{\p R}}_{M,Q}\ .
\en

The first law of thermodynamics applied to a system consisting of a thin shell characterized by $\paren{M,A,Q}$ is
\be
\dd S=\alpha\dd M+\alpha p \dd A-\alpha\Phi\dd Q\ .
\en
The extensive variables are implicit functions of $\paren{r_+,r_-,a}$, therefore we can rewrite the first law in the form
\be
\dd S=X\dd r_++Y\dd r_-+Z\dd a\ .
\en 
Using relations \eqref{eq:rel1}, \eqref{eq:rel2} and \eqref{eq:rel3}, $X$, $Y$ and $Z$ are given by
\begin{align}
X=\paren{\frac{\p S}{\p r_+}}_{r_-,a}&=\paren{\frac{\p S}{\p M}}_{A,Q}\paren{\frac{\p M}{\p r_+}}_{r_-,a}+\paren{\frac{\p S}{\p A}}_{M,Q}\paren{\frac{\p A}{\p r_+}}_{r_-,a}+\paren{\frac{\p S}{\p Q}}_{M,A}\paren{\frac{\p Q}{\p r_+}}_{r_-,a} \nonumber \\
&=\alpha\paren{\frac{\p M}{\p r_+}}_{r_-,a}+\alpha p\paren{\frac{\p A}{\p r_+}}_{r_-,a}-\alpha\Phi\paren{\frac{\p Q}{\p r_+}}_{r_-,a}\ , \nonumber \\
Y
&=\alpha\paren{\frac{\p M}{\p r_-}}_{r_+,a}+\alpha p\paren{\frac{\p A}{\p r_-}}_{r_+,a}-\alpha\Phi\paren{\frac{\p Q}{\p r_-}}_{r_+,a}\ , \nonumber \\
Z
&=\alpha\paren{\frac{\p M}{\p a}}_{r_+,r_-}+\alpha p\paren{\frac{\p A}{\p a}}_{r_+,r_-}-\alpha \Phi\paren{\frac{\p Q}{\p a}}_{r_+,r_-}\ , \nonumber
\end{align}
respectively. A straightforward calculation, after some algebra, yields
\begin{align}
X=\mathrm{a}\paren{1-\frac{\mathrm{c}r_-}{2}}\ , \\
Y=\mathrm{a}\paren{1-\frac{\mathrm{c}r_+}{2}}\ , \\
Z= 0\ .
\end{align}
Finally, the entropy differential can be written as
\be
\dd S=\mathrm{a}\paren{1-\frac{\mathrm{c}r_-}{2}}\dd r_++\mathrm{a}\paren{1-\frac{\mathrm{c}r_+}{2}}\dd r_-\ .
\en

\bibliography{RNthinshell.bib}

\end{document}